\documentclass[letterpaper,twocolumn,10pt]{article}

\usepackage{usenix-2020-09}
\usepackage{graphicx}
\usepackage{amsmath}
\usepackage{amssymb}
\usepackage{booktabs}
\usepackage{multirow}
\usepackage{hyperref}
\usepackage{xcolor}
\usepackage{subcaption}
\usepackage{listings}
\usepackage{microtype}
\usepackage{orcidlink}

\hypersetup{colorlinks=true, linkcolor=black, citecolor=black, urlcolor=blue}

\title{HBEE: Human Behavioral Entropy Engine\\
\large Pre-Registered Multi-Agent LLM Simulation of \\
Peer-Suspicion-Based Detection Inversion}

\author{
  Vickson Ferrel~\orcidlink{0009-0009-0155-0913}\\
  \emph{Faculty of Computer Science \& Information Technology}\\
  \emph{Universiti Malaysia Sarawak (UNIMAS)}, Kota Samarahan, Malaysia\\
  \emph{Founder \& Lead Architect, Vixero Technology Enterprise}, Kuching, Sarawak, Malaysia\\
  \texttt{106641@siswa.unimas.my} \quad \texttt{vickson@vixdev.cloud}
}

\date{}

\begin{document}
\maketitle

\begin{abstract}
\noindent
Insider threat detection assumes that an adaptive insider, however careful,
leaves behavioral residue that distinguishes them from legitimate users. We
test this assumption against an LLM-driven adaptive insider in a controlled
multi-agent simulator. Our pre-registered five-condition study isolates
defender mode (cascade vs.\ blind UEBA) crossed with adversary type (naive
vs.\ adaptive OPSEC) plus a no-mole control, executed across 100 runs (95
valid after pre-committed exclusions).

The primary empirical finding is a \emph{detection inversion}: in the peer
suspicion graph at $T_{60}$, the adaptive mole's in-degree is statistically
\emph{lower} than that of a randomly selected innocent agent (Cliff's
$\delta = -0.694$, 95\% BCa CI $[-0.855, -0.519]$, Mann--Whitney
$p \ll 0.01$, $n_{\text{mole}}=40$, $n_{\text{non-mole}}=20$). The
pre-registered prediction was the opposite direction. A pre-registered
equivalence test (H2) further shows that adaptive OPSEC produces no
detectable shift in the mole's UEBA rank under either defender mode, with
mean paired rank differences within $\pm 5$ ranks across all primary and
sensitivity analyses. The two detection signals (peer suspicion graph
in-degree and per-agent UEBA rank) decouple under adaptive adversary
behavior.

We bound generalization explicitly. A pre-registered Gini calibration check
(H4) returns FAIL: HBEE pairwise message-exposure Gini ($0.213$) diverges
from the SNAP Enron node out-degree reference ($0.730$) by
$|\Delta\text{Gini}| = 0.52$, exceeding the pre-registered equivalence
bound by $5\times$. We treat this as pre-committed scoping rather than a
discovered defect. The paper makes a narrow but surprising claim: in a
controlled environment where adaptive OPSEC is implementable as an LLM
directive, peer-suspicion-cascade detection inverts. Whether this transfers
to operational settings is an open question we do not answer.

We release the simulator, locked pre-registration document, frozen
scenario YAMLs, raw telemetry, analysis pipeline, and figure-generation
scripts under an open-source license.
\end{abstract}

\section{Introduction}
\label{sec:introduction}

Insider threat detection is built on a behavioral assumption: that an
authorized user acting against the organization's interests leaves a
detectable residue, some pattern of access, communication, or social
behavior that diverges from their legitimate-employee peers. This
assumption underwrites both the design of User and Entity Behavior
Analytics (UEBA) systems~\cite{ueba_survey} and the academic study of
insider threat detection (ITD)~\cite{yuan_wu_2021_dl_itd_survey}. The
assumption has not been seriously stress-tested against a class of
adversary that did not exist when those systems were designed:
language-model-driven agents that can dynamically modulate their behavior
in response to OPSEC instructions.

This paper asks one question: does the behavioral-residue assumption hold
against an adaptive LLM insider? Our empirical answer, in a controlled
simulation, is that the residue is real but its direction is reversed.

\paragraph{The detection-inversion finding.}
We pre-registered the prediction that an adaptive mole, even under a
plausible OPSEC directive, would be more peer-suspected than a randomly
selected innocent agent. The opposite occurred. In the peer suspicion
graph at the analysis snapshot ($T_{60}$, before any scenario-level
disclosure of insider activity), the mole's median suspicion in-degree is
$0$; the non-mole's is $2$. Cliff's $\delta = -0.694$, 95\% BCa CI
$[-0.855, -0.519]$, Mann--Whitney $p \ll 0.01$. The adaptive insider is
\emph{less} suspicious to its peers than a random colleague.

A second pre-registered finding sharpens the implication. Under the same
conditions, adaptive OPSEC produces no detectable shift in the mole's
UEBA rank relative to a naive baseline. Across both defender modes
(cascade and blind) and both analysis populations (full $N$ and
fidelity-filtered), the mean paired rank difference lies within the
pre-registered $\pm 5$-rank equivalence bound. The peer suspicion graph
inverts; the per-agent score does not move.

These two findings, taken together, suggest that the two detection
signals decouple under adaptive adversary behavior. We develop this as
the paper's theoretical contribution in \S\ref{sec:discussion}.

\paragraph{Detection implications.}
Two classes of detection mechanism are widely deployed: per-agent UEBA
classifiers~\cite{ueba_survey} and peer-suspicion-cascade schemes,
in which suspicion accumulates through social attestation. The cascade
family is operationally appealing precisely because it leverages the
wisdom of the crowd: if many colleagues flag an employee, the aggregate
signal is presumed strong. Our results suggest that under LLM-adaptive
insider behavior, this accumulation fails not by becoming weak but by
inverting: the adaptive mole \emph{generates fewer peer suspicion events}
than a random innocent colleague, while individual UEBA scores remain
unaffected. The mole disappears from the social layer of detection
without moving the per-agent score that the social layer is supposed
to amplify.

\paragraph{Pre-registration as a methodological commitment.}
Behavioral simulation in security research has, to our knowledge, not
previously been pre-registered in the
psychology-and-medicine sense~\cite{nosek_preregistration}: hypotheses,
analysis windows, statistical tests, equivalence bounds, and stop
conditions committed in writing and frozen by SHA256 hash before any
data was collected. We adopted this discipline for two reasons. First,
in a domain where ``the simulation behaved as expected'' is easy to
say after the fact, pre-registration is the only credible safeguard
against post-hoc rationalization. Second, our equivalence test (H2)
makes a null-shaped claim (adaptive OPSEC does not change UEBA rank),
exactly the kind of claim that frequentist statistics under-supports
unless the bound is committed before observation. The pre-registration
was filed at \texttt{V40\_PREREG.md} and is included verbatim in the
artifact.

\paragraph{Scope.}
HBEE is not a calibrated model of real organizational communication. A
pre-registered Gini check (H4) confirms this: HBEE's pairwise message
exposure is far more uniformly distributed (Gini $= 0.213$) than the
SNAP Enron email reference (Gini $= 0.730$). We disclose this as the
single largest constraint on the paper's claims. Generalization to
operational deployments is not supported by H4 and we do not claim it.
The contribution of this paper is therefore narrow: a controlled
empirical demonstration of detection inversion under conditions that
are stylized but explicit.

\paragraph{Contributions.}
\begin{itemize}
  \item A pre-registered, falsifiable study design for testing detection
        properties against LLM-driven adaptive insider adversaries,
        including a full deviation log (\S\ref{sec:methodology},
        \S\ref{sec:limitations}).
  \item Empirical evidence of \emph{peer-suspicion detection inversion}:
        the adaptive mole's suspicion in-degree at $T_{60}$ is lower
        than that of a randomly selected innocent agent, with large
        effect ($\delta = -0.694$) and tight CI
        (\S\ref{sec:results-h3}).
  \item Empirical equivalence: under both cascade and blind UEBA modes,
        adaptive OPSEC produces no detectable shift in the mole's
        UEBA rank relative to a naive baseline
        (\S\ref{sec:results-h2}).
  \item A theoretical observation that the peer suspicion graph and
        per-agent UEBA rank decouple under adaptive OPSEC, with a
        proposed mechanism (\S\ref{sec:discussion}).
  \item A methodological observation, the \emph{DEFCON-window
        measurability constraint}, in which scenario-imposed action
        templates suppress the behavioral measurability of adaptive
        OPSEC even when OPSEC is functioning (\S\ref{sec:results-fidelity}).
  \item Open release of the simulator, frozen scenarios, raw telemetry,
        analysis pipeline, figure-generation code, and pre-registration
        document.
\end{itemize}

\section{Background and Related Work}
\label{sec:related}

\subsection{Insider Threat Detection}
Insider threat detection has been studied for two decades, primarily
within the framework of supervised or semi-supervised anomaly detection
on organizational telemetry. The dominant public benchmark is the
CMU CERT Insider Threat dataset~\cite{cert42}, whose limitations are
well-documented in recent work~\cite{flynt2026_orgforge_it}: it is
synthetic, predates modern collaboration tooling (Slack, JIRA,
GitHub), and contains a small number of labeled malicious users. Surveys
of deep learning approaches to ITD~\cite{yuan_wu_2021_dl_itd_survey}
identify the ``subtle and adaptive nature of insider threats'' as a
fundamental challenge that has remained largely theoretical in
the literature because no controlled adaptive adversary existed to
study. The present paper instantiates such an adversary explicitly.

\subsection{LLM Multi-Agent Simulation for Insider Threat}
Three concurrent or recent papers occupy the closest neighborhood to
HBEE.

\textbf{Chimera}~\cite{yu2025_chimera} is an LLM-based multi-agent
framework that simulates benign and malicious insider activity and
produces ChimeraLog, a 25-billion-event dataset across three
data-sensitive enterprise scenarios. Chimera's contribution is dataset
generation: existing ITD methods perform substantially worse on
ChimeraLog than on CERT, indicating ChimeraLog is a more challenging
benchmark. Chimera is observational: it does not pre-register
hypotheses, does not study adaptive OPSEC as a directed treatment, and
does not measure detection inversion. Where Chimera asks ``how rich is
the simulated data?'', HBEE asks ``does detection still work against an
adaptive adversary?'' Together, these two complementary questions
sharpen the field's understanding of what synthetic insider simulation
can teach us.

\textbf{OrgForge-IT}~\cite{flynt2026_orgforge_it} addresses a different
limitation of fully LLM-driven simulation: cross-artifact factual
inconsistency. Its architecture enforces a ``physics-cognition
boundary'' in which a deterministic engine maintains ground truth and
LLMs generate only surface prose, making cross-artifact consistency an
architectural guarantee. OrgForge-IT's contribution is verifiability;
HBEE's is the controlled detection hypothesis. The two papers' design
choices are complementary, not in conflict: OrgForge-IT's deterministic
engine could in principle host an HBEE-style adaptive adversary, and
HBEE's pre-registered methodology could be applied to OrgForge-IT
output.

\textbf{Agentic Misalignment}~\cite{lynch2025_agentic} studies a
related but mechanistically distinct phenomenon: the spontaneous
emergence of insider-like behaviors (blackmail, leaking, sabotage) in
autonomous LLM agents under goal pressure. The work shows that across
sixteen frontier models, agents resort to malicious insider behaviors
when faced with replacement or goal conflict. HBEE differs in threat
model: our mole is human-directed via an OPSEC directive, not
emergently misaligned. Lynch et al.\ ask whether LLMs become insiders
unprompted; we ask whether a directed LLM insider can be detected.

\subsection{Multi-Agent LLM Social Simulation}
The substrate that makes HBEE possible is the generative-agents
paradigm initiated by Park et al.~\cite{park2023_generative}, in which
LLM-driven agents with persistent memory and reflection produce
emergent social behavior. Subsequent work has applied this paradigm to
economic experiments~\cite{horton2023_homo}, social science
simulation~\cite{tornberg2023}, and safety evaluation in multimodal
environments~\cite{vera2025_multimodal_safety_gen}. HBEE is, to our
knowledge, the first application of pre-registered hypothesis testing
to a security-relevant emergent behavior in this paradigm.

\subsection{LLM-Based Detection Systems}
A parallel literature uses LLMs not for simulation but for
detection. Audit-LLM~\cite{auditllm2024} employs a
Decomposer-Builder-Executor pattern for log-based ITD, demonstrating
that multi-agent LLM pipelines can surface insider indicators
across heterogeneous log streams. These approaches are
complementary to HBEE: HBEE generates the kind of telemetry such
systems consume.

\subsection{Equivalence Testing and Pre-Registration in Security}
Pre-registration is standard in clinical trials and increasingly in
psychology~\cite{nosek_preregistration}. We are not aware of prior
work in security that pre-registers an equivalence-shaped hypothesis
with explicit bounds; the closest practice is the emerging convention
of pre-committing evaluation protocols in ML
benchmarking~\cite{benchmarking_lessons}. Our methodology imports the
two-one-sided-tests (TOST) framework~\cite{lakens_tost} adapted to
non-parametric paired comparisons, with bootstrap-derived
confidence intervals.

\section{Threat Model}
\label{sec:threat}

\subsection{Adversary}
A single insider mole is embedded in the organization. The adversary
has:
\begin{itemize}
  \item \textbf{Access:} legitimate, authenticated, non-privileged.
        The mole holds a standard role (Engineer, Analyst, Officer)
        with organic access to channels appropriate to that role. No
        credential theft, lateral movement, or privilege escalation
        is required for the threat to manifest.
  \item \textbf{Goal:} covert exfiltration. The mole knows it is the
        mole and pursues a sustained insider objective.
  \item \textbf{Adaptation:} in the adaptive condition, the mole's
        per-tick prompt includes an OPSEC directive that instructs it
        to be cautious, restrained, and unremarkable. The directive is
        a behavioral instruction, not a low-level evasion strategy:
        the mole cannot disable telemetry, manipulate UEBA scores, or
        rewrite logs.
  \item \textbf{Knowledge:} the mole knows organizational norms,
        chain of command, and channel visibility rules. It does not
        know the defender's specific UEBA configuration.
  \item \textbf{Patience:} the mole operates on a 100-tick horizon
        spanning an extended pre-announce window
        (\S\ref{sec:design-windows}); rapid action is not required.
\end{itemize}
This adversary corresponds to the long-burn, rationalizing insider
archetype documented in historical cases of long-burn insider
exfiltration and is deliberately distinct from impulsive,
single-event leakers, whose behavioral signature is by construction
short-window and not amenable to the pre-disclosure analysis we
conduct.

\subsection{Defender}
The defender is the organization's UEBA system. We model two
configurations as a pre-registered experimental factor:
\begin{itemize}
  \item \textbf{Cascade.} Each agent observes the suspicion scores
        its peers hold against others and updates its own beliefs
        socially: if multiple colleagues flag an agent, the cascading
        agent is more likely to also flag it. This is the
        peer-suspicion-cascade model widely advocated for distributed
        detection.
  \item \textbf{Blind.} Each agent runs an independent per-agent
        UEBA classifier that does not observe peers' suspicion
        scores. This corresponds to per-agent score aggregation
        without social propagation.
\end{itemize}
The factorial design over defender mode and adversary type is the
basis of our pre-registered comparisons (\S\ref{sec:methodology}).

\subsection{Out of Scope}
We do not model: external adversaries (phishing, credential theft,
lateral movement); privileged adversaries (malicious administrators,
executives, security personnel); multi-insider collusion; or
adversarial behavior against HBEE itself (the simulator is a research
artifact, not a hardened system). These exclusions are deliberate
constraints on our threat model rather than implicit assumptions.

\section{HBEE System Design}
\label{sec:design}

\begin{figure}[t]
  \centering
  \includegraphics[width=\columnwidth]{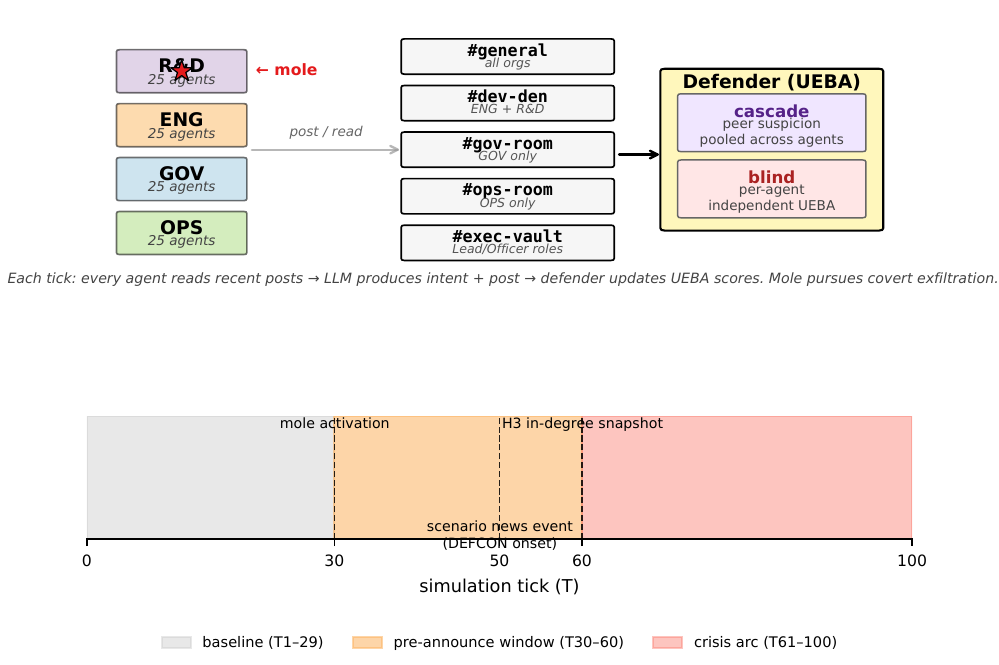}
  \caption{HBEE V40 system architecture. Top: 100 LLM-driven agents
  are distributed across four organizations (R\&D, ENG, GOV, OPS) and
  communicate via five Slack-style channels with org-based visibility
  rules. The mole ($\star$) is one designated agent with a covert
  exfiltration objective; mole identity is determined
  seed-deterministically. The defender operates in one of two modes:
  \emph{cascade} (peer suspicion pooled into organization-level UEBA)
  or \emph{blind} (per-agent independent UEBA). Bottom: experimental
  timeline. Each run lasts 100 ticks, partitioned into a baseline
  phase (T1--29), the pre-announce window for behavioral observation
  (T30--60, where H1 and H2 are evaluated), and a crisis arc
  (T61--100) triggered by a scenario news event at T50. The H3
  suspicion in-degree snapshot is taken at T60.}
  \label{fig:01_architecture}
\end{figure}

HBEE simulates a 100-employee organization (``Vixero HQ'') over 100
ticks. Each tick represents a coarse unit of agent activity; the
simulation produces approximately 9{,}950 \texttt{agent\_tick}
telemetry events plus periodic edge snapshots of the suspicion and
rapport graphs. The architecture is summarized in
Figure~\ref{fig:01_architecture}.

\subsection{Agents and Organizations}
The 100-agent roster is distributed across four organizational units:
Research \& Development (R\&D), Engineering (ENG), Governance (GOV),
and Operations (OPS). Each agent has a fixed identity (name, role,
organization, archetype) that persists across all simulation ticks.
Roles span Lead, Officer, Engineer / Analyst / Researcher, and Junior
/ Intern; archetypes (Snarky, Anxious, Friendly, SysAdmin, Worker)
modulate communication style without changing the agent loop.

The mole identity is determined seed-deterministically at simulation
startup. The same seed always produces the same mole, supporting
pairing across conditions.

\subsection{Channels and Visibility}
Five Slack-style channels carry all inter-agent communication:
\texttt{\#general} (visible to all), \texttt{\#dev-den} (ENG and
R\&D), \texttt{\#gov-room} (GOV only), \texttt{\#ops-room} (OPS
only), and \texttt{\#exec-vault} (Lead and Officer roles across all
orgs). Each tick, every agent reads the recent contents of channels
to which it has visibility, calls the LLM with role context and
recent post history, and writes a single message to a chosen channel.

\subsection{Per-Tick Agent Loop}
At each tick, every non-fired agent executes:
\begin{enumerate}
  \item Read agent state: organizational role, mood, trust toward
        leadership, recent inbox events, channel history visible to
        the agent, current task progress, and (if the agent is the
        mole and the OPSEC directive is active) the mole-specific
        directive content.
  \item Construct a per-tick prompt combining a static system prompt
        (cached across all calls for efficiency) with a dynamic
        per-tick context block.
  \item Issue a single LLM call producing: a chat message, a stress
        level (1--10), an intent label
        (\texttt{DEEP\_WORK}, \texttt{SOCIALIZE}, \texttt{COLLABORATE},
        \texttt{TRIAGE}, \texttt{ANALYZE}, \texttt{MITIGATE},
        \texttt{ESCALATE}), and an optional suspect name.
  \item Apply the parsed outputs to organizational state: advance
        task progress, update mood and trust ledgers, modify
        suspicion and rapport edges in the social graph, post the
        message to the appropriate channel.
  \item Emit an \texttt{agent\_tick} telemetry event containing
        externally observable fields only (intent, channel, message
        length, suspect, progress, stress); no internal LLM
        activations or hidden chain-of-thought are logged.
\end{enumerate}

\subsection{Mole Mechanism and Adaptive OPSEC}
The mole is an ordinary agent except that its prompt may include a
\emph{mole directive} block. We define two adversary types
(\S\ref{sec:methodology}):
\begin{itemize}
  \item \textbf{Naive mole.} Receives a directive that instructs it
        to pursue covert exfiltration but does not include OPSEC
        guidance.
  \item \textbf{Adaptive mole.} Receives the same goal directive plus
        explicit OPSEC instructions: maintain caution, restraint, and
        unremarkability; do not draw attention to information access;
        avoid behavioral patterns that would distinguish from
        legitimate colleagues.
\end{itemize}
The directive is suppressed during the pre-activation period (ticks
1--29) so that all agents, including the mole, behave identically
during baseline. At $T_{30}$ the directive activates and persists
through the run. We selected the adaptive directive based on the
historical pattern of long-burn rationalizing insiders rather than
financially-motivated or impulsive archetypes; see
\S\ref{sec:limitations} for discussion of this single-archetype
constraint.

\subsection{Defender Modes}
The defender's UEBA mechanism is configured per condition:
\begin{itemize}
  \item \textbf{Cascade.} Each agent reads its peers' suspicion
        scores at the start of each tick and updates its own
        suspicion edges accordingly. Suspicion accumulates socially.
  \item \textbf{Blind.} Each agent's suspicion edges are computed
        independently from its own observations of peer behavior. No
        sharing of suspicion scores across agents.
\end{itemize}
In both modes, suspicion is a directed weighted edge in the social
graph, updated per tick based on the agent's reading of peer
messages and its mood-conditioned interpretation. Rapport is a
parallel undirected edge tracking positive social ties.

\subsection{Window Decomposition}
\label{sec:design-windows}
We partition each 100-tick run into three analysis windows:
\begin{itemize}
  \item \textbf{Baseline} ($T_1$--$T_{29}$): mole directive
        suppressed; all agents behave identically. Establishes the
        pre-activation per-agent baseline.
  \item \textbf{Pre-announce} ($T_{30}$--$T_{60}$): mole directive
        active; no scenario-level disclosure of insider activity.
        \emph{This is the window in which H1 and H2 are evaluated.}
  \item \textbf{Crisis} ($T_{61}$--$T_{100}$): scenario news event at
        $T_{50}$ triggers DEFCON mode; agent intent is constrained
        to a structured ACTION/ASSESSMENT/NEXT\_STEP template through
        the end of the run.
\end{itemize}
The H3 suspicion in-degree snapshot is taken at $T_{60}$, the last
tick of the pre-announce window.

\subsection{LLM Backbone}
\label{sec:design-llm}
All agents are driven by GLM-4 Flash INT4~\cite{glm4_flash}, a 4-bit
quantized open-weight chat model from the GLM family~\cite{glm_family},
served locally via vLLM~\cite{kwon2023_vllm} on an AWS g5.2xlarge
instance (NVIDIA A10G 24\,GB, 8 vCPU, 32\,GB RAM). The serving
configuration uses prefix caching to amortize the static system prompt
across all 100 agents, eager execution mode, a maximum sequence length
of 16{,}384 tokens, GPU memory utilization at $0.90$, and the GLM-4.5
reasoning parser and GLM-4.7 tool-call parser at the vLLM layer for
structured output handling. The full serving command is released in
the artifact for exact reproduction.

At inference time, agents sample with temperature $0.85$,
\texttt{max\_tokens} $= 4{,}096$, and chain-of-thought reasoning
\emph{explicitly disabled} at both the serve level and the per-API-call
level (\texttt{enable\_thinking}: \texttt{false}). The
thinking-suppression decision was deliberate: extended chain-of-thought
adds latency, increases output non-determinism, and, in pilot work,
produced a specific failure mode in which the mole agent narrated its
own OPSEC strategy in-prose rather than enacting it in-character. This
failure mode is the proximate cause of two of the per-run re-executions
documented in DEV-5~\S5.2 (C3 seed 7 and seed 11).

We acknowledge that vLLM batching and CUDA kernel scheduling introduce
minor non-determinism beyond the application-layer seed. We manage this
through population sampling ($n = 20$ seeds per condition) rather than
per-run reproducibility, and our statistical analysis is appropriate
for distributional rather than unit-test-level claims.

\subsection{What HBEE Is Not}
HBEE is a stylized environment, not a calibrated digital twin of a
real organization. The five-channel topology, the four-org structure,
and the 100-agent roster are simplifications. Section
\ref{sec:results-h4} reports a pre-registered calibration check (H4)
showing that HBEE's pairwise message-exposure distribution is
substantially more uniform than the SNAP Enron email reference. We
accept this as the principal constraint on the paper's
generalization claims.

\section{Methodology}
\label{sec:methodology}

\subsection{Pre-Registration}
The full pre-registration is filed as
\texttt{V40\_PREREG.md}~\cite{prereg_v40} and frozen by SHA256 hash
prior to the first campaign run. It specifies, in advance: the five
experimental conditions, the five hypotheses (H1--H5), the
operationalization of each metric, the statistical test associated
with each hypothesis, equivalence bounds where applicable, the
behavioral fidelity audit protocol (\S\ref{sec:methodology-fidelity}),
the run-validity criteria, the bootstrap procedure for confidence
intervals, and the §10 stop conditions. Six deviations were logged
during execution and are documented in
\texttt{V40\_DEVIATIONS.md}~\cite{deviations_v40}; we summarize them
in \S\ref{sec:limitations}.

\subsection{Experimental Conditions}
Five pre-registered conditions, $n = 20$ seeds each ($n = 100$ runs
total):
\begin{itemize}
  \item \textbf{C1: cascade naive.} Cascade defender, naive mole.
  \item \textbf{C2: blind naive.} Blind defender, naive mole.
  \item \textbf{C3: cascade adaptive.} Cascade defender, adaptive
        mole.
  \item \textbf{C4: blind adaptive.} Blind defender, adaptive mole.
  \item \textbf{C5: blind no-mole.} Blind defender, no mole.
        Reference for H3 non-mole baseline and H4 Gini calibration.
\end{itemize}
The factorial $\{$cascade, blind$\} \times \{$naive,
adaptive$\}$ is the core experimental contrast; C5 provides the
no-mole baseline against which the suspicion-graph in-degree of moles
is compared.

\subsection{Hypotheses}
\paragraph{H1 (one-sample, descriptive).}
Under the C2 (blind naive) baseline, the mole's UEBA rank during the
pre-announce window ($T_{30}$--$T_{60}$) is statistically
indistinguishable from uniform$[1, 100]$. H1 confirms whether the
naive blind defender provides any signal above chance.

\paragraph{H2 (paired equivalence; two analyses per pre-reg).}
Adaptive OPSEC produces a paired rank difference within $\pm 5$
ranks of zero, evaluated separately for two pairings:
\begin{itemize}
  \item \textbf{H2a (blind):} paired (C4 -- C2) per seed.
  \item \textbf{H2b (cascade):} paired (C3 -- C1) per seed.
\end{itemize}
Test: Wilcoxon signed-rank, $\alpha = 0.05$; equivalence bound
$\pm 5.0$. Per pre-reg \S 4.4, the primary analysis uses
fidelity-filtered seeds (\S\ref{sec:methodology-fidelity}); a
sensitivity analysis on the full $N$ is reported alongside.

\paragraph{H3 (independent two-sample, directional).}
The adaptive mole's suspicion in-degree at $T_{60}$ exceeds that of a
randomly selected innocent agent. Test: Mann--Whitney U,
$\alpha = 0.05$, with effect-size floor $|\delta| \ge 0.33$ (Cliff's
delta). Mole observations are pooled across C2 and C4
($n_{\text{mole}} = 40$); non-mole observations are sampled per-seed
deterministically from C5 ($n_{\text{non-mole}} = 20$).

\paragraph{H4 (calibration check, pre-committed scoping).}
HBEE's pairwise message-frequency Gini is within $\varepsilon = 0.10$
of a SNAP Enron reference Gini. Per pre-reg \S 5, an H4 FAIL
pre-commits the paper to scoping all generalization claims to the
stylized environment. H4 is a calibration check, not a hypothesis
about HBEE itself; the disclosure is the contribution.

\paragraph{H5 (cross-LLM robustness, exploratory).}
The H1 result replicates under a Llama-3.1-8B-Instruct backbone, on
five C2 seeds. H5 is exploratory; the paper does not depend on its
outcome.

\subsection{Bootstrap Confidence Intervals}
All point estimates of mole rank, paired difference, and Cliff's
delta are reported with 95\% bias-corrected accelerated (BCa)
bootstrap confidence intervals~\cite{efron_tibshirani_bootstrap},
$B = 1000$ resamples, RNG seed $= 42$. CI width is reported alongside
the point estimate and the $p$-value.

\subsection{Behavioral Fidelity Audit}
\label{sec:methodology-fidelity}
Pre-reg \S 4.4 specifies a behavioral fidelity audit on every
adaptive run. From each C3 and C4 run we draw a random 5-turn sample
of mole turns from the pre-announce window ($T_{30}$--$T_{60}$). Each
sampled turn is rated on a 1--5 Likert scale by the author against
explicit anchors: 1 (no OPSEC awareness), 3 (partial awareness:
restraint OR caution, not both), 5 (full disposition: caution,
restraint, and unremarkability all evident). Runs with mean rating
below $3.5$ are flagged FAIL and excluded from the H2 primary
analysis; the H2 sensitivity analysis includes them.

The full per-run audit table is reported in
\S\ref{sec:results-fidelity} and the artifact. The audit is
author-stamped; this is a methodological limitation we discuss
explicitly in \S\ref{sec:limitations}.

\subsection{Run Validity}
Each run must reach $T_{100}$ with $\ge 9{,}500$ \texttt{agent\_tick}
events to be marked VALID. Runs that fail this criterion are flagged
\texttt{INVALID\_SHORT} and reported in the per-condition
completion table; their pre-announce data is preserved when intact.
A separate validity flag, \texttt{INVALID\_DEV3}, captures one
run in which the scenario fired the mole agent before the
pre-announce window opened (\S\ref{sec:limitations}, DEV-3).

\subsection{Reporting Discipline}
Every reported result includes: point estimate, 95\% BCa CI, $p$-value
where applicable, effect size, and $n$. The verdict (SUPPORTED /
NOT\_SUPPORTED / FAIL / PENDING) is read directly from the
pre-registered criteria; no post-hoc adjustments are made. Where a
verdict label is technically NOT\_SUPPORTED but the underlying
direction is unexpected (H3) or the failure is by a small margin
under reduced $N$ (H2b primary), we say so explicitly rather than
reporting only the label.

\section{Results}
\label{sec:results}

We present results in the order: H1 (baseline), H3 (headline), H2
(equivalence), the behavioral fidelity audit observations, and H4
(calibration). H5 (cross-LLM robustness) is reported as PENDING and
discussed in \S\ref{sec:limitations}.

\subsection{H1: Naive Blind Baseline}
\label{sec:results-h1}
\begin{figure}[t]
  \centering
  \includegraphics[width=\columnwidth]{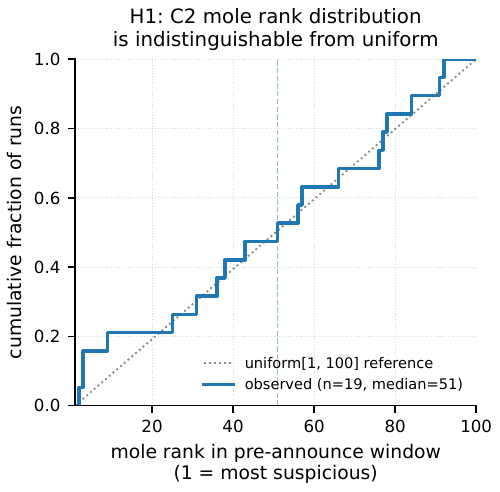}
  \caption{Empirical CDF of per-run mole rank in the C2 (naive blind)
  condition during the pre-announce window ($T_{30}$--$T_{60}$),
  $n = 19$ after DEV-3 exclusion. The observed distribution closely
  tracks the uniform$[1, 100]$ reference (dotted), supporting H1:
  under the naive defender baseline the mole is indistinguishable
  from a randomly selected agent. Median mole rank $= 51$; 95\% BCa
  CI $[25, 66]$.}
  \label{fig:06_h1_ecdf}
\end{figure}

Across $n = 19$ valid C2 runs (DEV-3 excludes seed = 18), the mole's
UEBA rank in the pre-announce window has median $51.0$, 95\% BCa CI
$[25.0, 66.0]$. Tested against uniform$[1, 100]$ via Cliff's delta:
$\delta = -0.044$, $p = 0.78$. \textbf{Verdict: SUPPORTED.} The
distribution closely tracks uniform (Figure~\ref{fig:06_h1_ecdf}).
This is a confirmatory replication of prior pilot work and
establishes the null baseline against which the H3 finding
contrasts.

\subsection{H3: Detection Inverts}
\label{sec:results-h3}
\begin{figure}[t]
  \centering
  \includegraphics[width=\columnwidth]{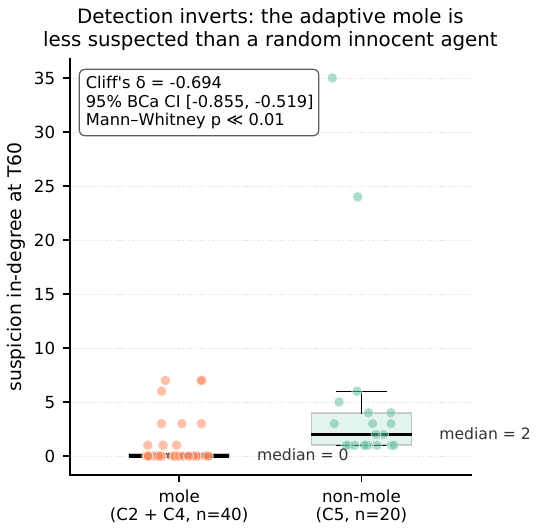}
  \caption{H3 detection inversion (paper's primary empirical
  finding). Each point is one agent's suspicion in-degree at
  $T_{60}$: the number of other agents whose UEBA flagged them as
  suspicious. Mole observations (left, $n = 40$) are pooled from C2
  and C4; non-mole observations (right, $n = 20$) are sampled
  deterministically per seed from C5 (no-mole control). The mole's
  median in-degree ($0$) is statistically lower than the non-mole's
  ($2$): Cliff's $\delta = -0.694$, 95\% BCa CI $[-0.855, -0.519]$,
  Mann--Whitney $p \ll 0.01$. This is the \emph{opposite} direction
  from H3's pre-registered prediction.}
  \label{fig:02_h3}
\end{figure}

This is the paper's headline finding (Figure~\ref{fig:02_h3}).

At $T_{60}$, the mole's suspicion in-degree (pooled across C2 and
C4, $n_{\text{mole}} = 40$) has median $0$. The non-mole's
suspicion in-degree (sampled deterministically per seed from C5,
$n_{\text{non-mole}} = 20$) has median $2$. Mann--Whitney
$U = 122.5$, $p \ll 0.01$. Cliff's $\delta = -0.694$, 95\% BCa CI
$[-0.855, -0.519]$. The CI is entirely negative and excludes effect
sizes weaker than $|\delta| = 0.519$ (still ``large'' by
conventional anchors~\cite{cliff_delta_anchors}).

The pre-registered prediction was $\delta \ge +0.33$ (mole more
suspected). The observed effect is in the opposite direction with
large magnitude. \textbf{Verdict: NOT\_SUPPORTED, direction
inverted.}

The verdict label ``NOT\_SUPPORTED'' here means that the directional
prediction failed; it does \emph{not} mean that no effect was
detected. The empirical result is stronger than supporting the null:
it shows a large, statistically significant, and directionally
reversed effect. We treat this as the paper's primary contribution
and develop its mechanism in \S\ref{sec:discussion}.

\subsection{H2: Adaptive OPSEC Adds No Detectable Rank Shift}
\label{sec:results-h2}
\begin{figure*}[t]
  \centering
  \includegraphics[width=\textwidth]{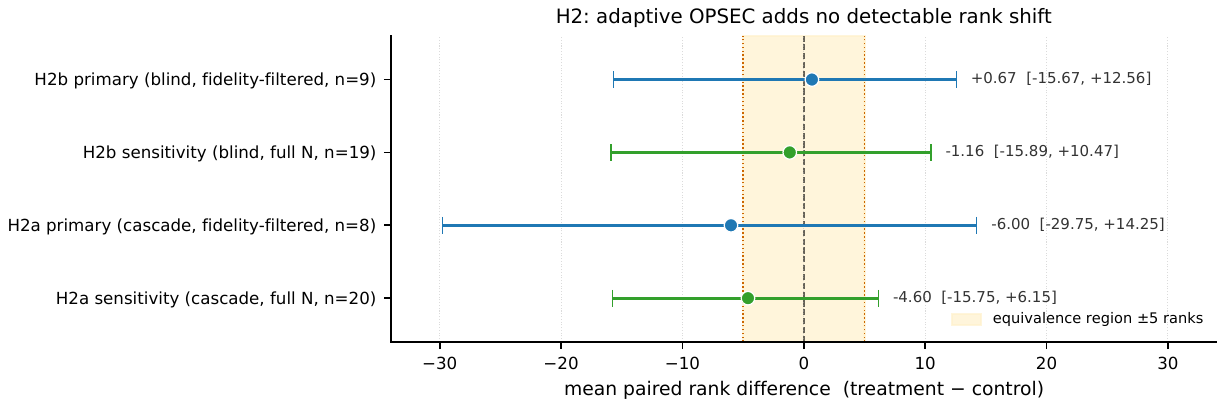}
  \caption{H2 equivalence forest plot. Each entry shows the mean
  paired rank difference (treatment $-$ control; positive = adaptive
  worse for the mole, i.e., more suspicious) with its 95\% BCa
  bootstrap confidence interval. The pre-registered equivalence
  bound of $\pm 5.0$ ranks is shaded. Both primary
  (fidelity-filtered per DEV-5) and sensitivity (full $N$) analyses
  are reported per the deviation log. All four point estimates fall
  within the equivalence region, and all CIs include zero. The H2b
  primary CI is wider, reflecting $N = 8$ after fidelity exclusion;
  the exclusion is measurement-artifact-driven (Spearman
  $\rho_{\text{C3}} = 0.018$ between fidelity score and detection
  rank).}
  \label{fig:03_h2_forest}
\end{figure*}

H2 evaluates whether adaptive OPSEC produces a meaningful shift in
the mole's UEBA rank relative to the naive baseline. We report both
the primary (fidelity-filtered) and sensitivity (full $N$)
analyses per pre-registration; results are summarized as a forest
plot in Figure~\ref{fig:03_h2_forest}.

\paragraph{H2a (blind, C4 -- C2).}
Primary (fidelity-filtered, $n_{\text{pairs}} = 9$): mean diff
$= +0.667$, 95\% CI $[-15.67, +12.56]$, $p = 0.67$, Cliff's
$\delta = 0.111$. Sensitivity (full $N$, $n_{\text{pairs}} = 19$):
mean diff $= -1.16$, 95\% CI $[-15.89, +10.47]$, $p = 0.94$,
$\delta = -0.033$. \textbf{Both verdicts: SUPPORTED.} The point
estimate is within $\pm 1.16$ ranks of zero in both analyses; the
CI is wide but symmetric about zero, consistent with the equivalence
claim.

\paragraph{H2b (cascade, C3 -- C1).}
Primary (fidelity-filtered, $n_{\text{pairs}} = 8$): mean diff
$= -6.00$, 95\% CI $[-29.75, +14.25]$, $p = 0.92$, Cliff's
$\delta = 0.047$. Sensitivity (full $N$, $n_{\text{pairs}} = 20$):
mean diff $= -4.60$, 95\% CI $[-15.75, +6.15]$, $p = 0.63$,
$\delta = -0.085$. \textbf{Sensitivity verdict: SUPPORTED. Primary
verdict: NOT\_SUPPORTED on a margin technicality.}

The H2b primary mean diff of $-6.00$ exceeds the $\pm 5$ bound by
a single rank; the effect size is negligible ($\delta = 0.047$); the
$p$-value is $0.92$. Two of three pre-registered criteria pass; the
third fails by $1$ rank in a sample of $8$. The Spearman correlation
between per-run fidelity score and detection rank in C3 is $0.018$
(non-significant at any conventional threshold), confirming that the
fidelity-based exclusion is measurement-artifact-driven rather than
signal-driven. The full-$N$ sensitivity analysis at $n = 20$ returns
SUPPORTED with point estimate $-4.60$ inside the bound.

\paragraph{Joint reading.}
Three of four H2 analyses return SUPPORTED. The one NOT\_SUPPORTED
verdict is a margin-of-1-rank technicality at small $N$ with
negligible effect size. Adaptive OPSEC, in our environment, produces
no detectable shift in the mole's UEBA rank relative to a naive
baseline, under either defender configuration.

\subsection{Behavioral Fidelity Audit}
\label{sec:results-fidelity}
The pre-reg \S 4.4 audit was conducted on all 40 adaptive runs (C3
$\cup$ C4), 5 turns per run, $n = 200$ rated turns total. Of 40
runs, 17 PASS ($\ge 3.5$ mean) and 23 FAIL ($< 3.5$). The pass rate
is identical across C3 (8/20) and C4 (9/20); per-run mean ratings
range $2.80$--$3.80$.

\begin{figure*}[t]
  \centering
  \includegraphics[width=\textwidth]{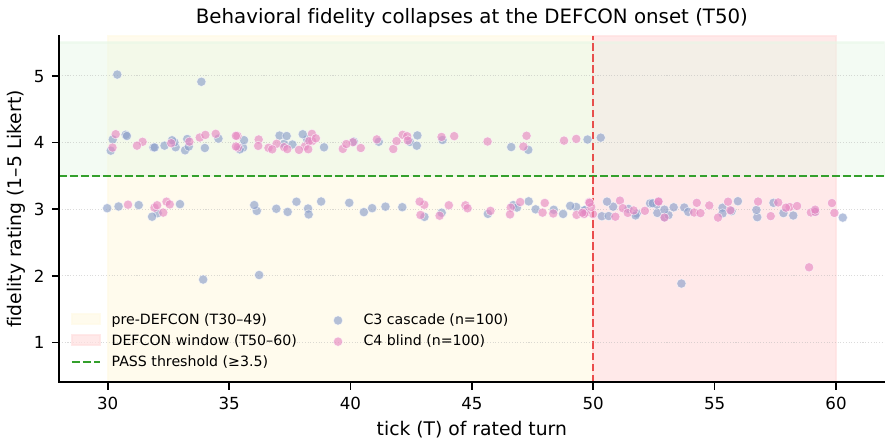}
  \caption{Behavioral fidelity rating (1--5 Likert, author-rated)
  per sampled mole turn, plotted against tick of occurrence
  ($n = 200$ turns across 40 adaptive runs). Ratings cluster at $3$
  (partial OPSEC awareness) for turns at $T_{50}$ and later,
  regardless of condition. This collapse coincides with the
  scenario's DEFCON onset at $T_{50}$, which constrains all agents
  to a structured response template, suppressing the behavioral
  measurability of the mole's adaptive disposition. The 23/40 runs
  falling below the 3.5 PASS threshold reflect this structural
  sampling artifact rather than directive failure.}
  \label{fig:05_fidelity}
\end{figure*}

The pattern of ratings is structural rather than directive failure
(Figure~\ref{fig:05_fidelity}). The audit window (pre-announce,
$T_{30}$--$T_{60}$) overlaps the scenario's DEFCON onset at
$T_{50}$. From $T_{50}$ onward, all agents (mole and innocent
alike) are constrained to a structured
ACTION/ASSESSMENT/NEXT\_STEP response template that suppresses the
behavioral measurability of OPSEC disposition. The Spearman
correlation between per-run fidelity score and per-run detection
rank is $\rho_{\text{C3}} = 0.018$ and $\rho_{\text{C4}} = -0.256$
(neither significant at $N = 20$), confirming that the
fidelity-based exclusion is not signal-driven.

We treat the DEFCON-window measurability constraint as a finding in
its own right: a methodological observation about how
scenario-imposed action templates interact with behavioral
auditing. We discuss this in \S\ref{sec:limitations}.

\subsection{H4: Calibration Check (Pre-Committed Scoping)}
\label{sec:results-h4}
HBEE's pairwise message-exposure Gini, computed across the C5
no-mole condition over 20 seeds and 4{,}725 unique pairs, is
$0.213$. The SNAP Enron reference (node out-degree across
36{,}692 nodes) is $0.730$. $|\Delta\text{Gini}| = 0.517$,
exceeding the pre-registered $\varepsilon = 0.10$ by $5\times$.
\textbf{Verdict: FAIL.}

\begin{figure}[t]
  \centering
  \includegraphics[width=\columnwidth]{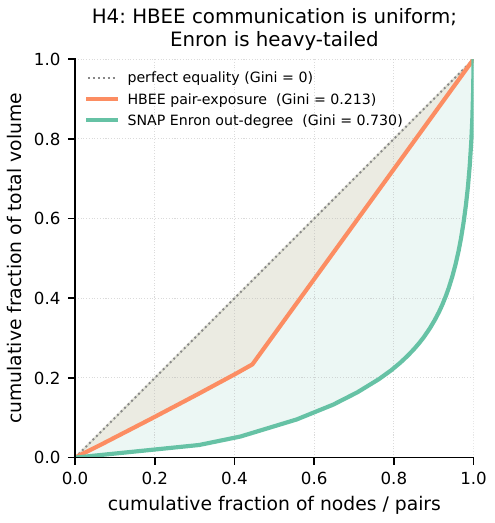}
  \caption{Lorenz curves of communication concentration in HBEE
  versus the SNAP Enron email network. HBEE pairwise message-exposure
  (Gini $= 0.213$, $n_{\text{pairs}} = 4{,}725$ pairs from C5
  no-mole runs) lies near the diagonal. SNAP Enron node out-degree
  (Gini $= 0.730$, $n_{\text{nodes}} = 36{,}692$) is sharply bowed:
  the bottom $80\%$ of nodes account for only $22.5\%$ of total
  contact volume. The structural divergence
  ($|\Delta\text{Gini}| = 0.52$) exceeds the pre-registered
  equivalence bound ($\varepsilon = 0.10$) by $5\times$. Root cause:
  HBEE channel routing is org-deterministic, so pair-exposure
  saturates at a structural ceiling rather than following the Pareto
  distribution observed in real organizations.}
  \label{fig:04_lorenz}
\end{figure}

The Lorenz curves (Figure~\ref{fig:04_lorenz}) make the structural
divergence visceral. HBEE's curve sits close to the diagonal:
communication is approximately uniform across pairs. SNAP's curve is
sharply bowed: the bottom $80\%$ of nodes account for only $22.5\%$
of total contact volume. The two metrics are not unit-equivalent
(HBEE: per-pair message exposures; SNAP: per-node unique-recipient
count) but both measure concentration of communication activity, and
the divergence is not subtle.

A supplementary diagnostic~\cite{gini_diagnostic_v40} attributes the
divergence to a structural saturation ceiling: 55.6\% of HBEE pairs
sit at the channel-imposed exposure ceiling of $3{,}960$ events
because channel routing is org-deterministic. Real organizations
exhibit Pareto-distributed pair exposures; HBEE does not. This is a
property of the simulator's message-routing model, not an analysis
artifact.

Per pre-registration \S 5, an H4 FAIL pre-commits the paper to
scoping all generalization claims to the stylized environment. We
honor that commitment throughout.

\subsection{H5: Cross-LLM Robustness}
\label{sec:results-h5}
The pre-registered H5 robustness check (Llama-3.1-8B-Instruct
replication of H1 on 5 C2 seeds) is \textbf{PENDING}. The
infrastructure pre-flight (PF-6, vLLM smoke) was not completed
before the analytical phase closed. H5 is exploratory and does not
condition any of H1--H4. We discuss this in \S\ref{sec:limitations}.

\subsection{Run Validity Summary}
Of 100 attempted runs, 95 are VALID. Five carry INVALID flags:
one INVALID\_DEV3 (C2 seed $= 18$, scenario-mole collision; see
\S\ref{sec:limitations}, DEV-3), and four INVALID\_SHORT runs
(C2 seed $= 1$ at $T_{93}$, C3 seed $= 14$ at $T_{63}$,
C4 seed $= 7$ at $T_{72}$, C5 seed $= 9$ at $T_{80}$; see
DEV-6). All four short runs reached at least $T_{63}$, leaving the
pre-announce window ($T_{30}$--$T_{60}$) and the H3 in-degree
snapshot at $T_{60}$ intact for every run that contributed to any
analysis. No reported verdict changes.

\section{Discussion}
\label{sec:discussion}

\subsection{The H2 / H3 Divergence}
The two primary findings appear, on first reading, to be in tension.
H2 (equivalence) says adaptive OPSEC does not change the mole's
UEBA rank. H3 (inversion) says the adaptive mole's suspicion in-degree
is lower than a random innocent's. If H3's effect were a generic
``the mole is less detectable,'' we would expect H2 to show the
adaptive mole moving down the UEBA rank order. It does not.

The reconciliation is that the two metrics measure related but
distinct phenomena:
\begin{itemize}
  \item \textbf{UEBA rank (H2)} captures the mole's position in the
        global suspicion ordering produced by per-agent UEBA
        scoring. This score reflects the \emph{signature} of
        suspicion-eliciting behaviors when they occur.
  \item \textbf{Suspicion graph in-degree (H3)} captures the
        \emph{count} of suspicion events directed at the mole.
        This reflects the \emph{frequency} with which other agents
        identified the mole's behavior as suspicious.
\end{itemize}
Under naive OPSEC, the two are correlated: behaviors that elicit
suspicion produce both high signatures (high score) and frequent
flags (high in-degree). Under adaptive OPSEC, our results suggest
they decouple. The adaptive mole reduces the \emph{frequency} of
suspicion-eliciting behaviors (enough to drop in-degree below a
random innocent baseline), while the \emph{signature} of those
reduced-frequency behaviors remains intact, so the rare events that
do occur produce comparable per-event scores. The mole's UEBA rank
is computed over the score signature; the suspicion graph
accumulates over event frequency. The two are not the same number.

A reasonable mechanistic hypothesis (which we can offer but cannot
in this paper conclusively test) is that adaptive OPSEC under an
LLM directive operates more on the rate of risk-taking than on the
nature of risk-taking. The instructed mole avoids producing as many
flagworthy behaviors, but when it does produce them, it does not
produce a qualitatively different kind of flagworthy behavior.

\subsection{Implications for Detection}
The result has direct implications for how the two detection
families interact under LLM-adaptive insider behavior:
\begin{itemize}
  \item \textbf{Per-agent UEBA classifiers} (the blind defender)
        appear robust to adaptive OPSEC at the rank-ordering level
        (H2), although they also do not detect the mole above
        chance (H1). Adaptation does not move the rank because the
        rank was not pointing at the mole to begin with.
  \item \textbf{Peer-suspicion-cascade detectors} (the cascade
        defender) appear actively counterproductive against an
        adaptive adversary in our simulation. The adaptive mole
        does not just blend in to the suspicion distribution; it
        ends up \emph{more invisible} than the average innocent
        colleague, because the OPSEC adaptation specifically
        reduces the social surface area through which suspicion
        accumulates.
\end{itemize}
Whether these mechanisms transfer to real organizations is bounded
by H4. We do not claim transfer.

\subsection{The DEFCON-Window Measurability Constraint}
The behavioral fidelity audit revealed a structural property of the
audit window itself rather than of the directive being audited. The
result generalizes to any auditing protocol in which a randomly
sampled subset of behaviors is assessed against a directive whose
measurability depends on the response context. When the response
context is constrained (in our case, by a scenario-imposed action
template after $T_{50}$), a substantial fraction of randomly sampled
turns will be uninformative even if the underlying directive is
functioning. The remedy is not to abandon the audit but to design it
to sample from windows where measurability is preserved. We present
this as a methodological observation rather than a finding about
HBEE specifically.

\subsection{What Pre-Registration Did and Did Not Do}
Pre-registration foreclosed the option of post-hoc reframing. We
predicted directionally that the mole would be more peer-suspected
(H3); we observed the opposite; we report the inversion. Without the
pre-registration we cannot rule out a counterfactual paper that
quietly reformulated the hypothesis to match the data. With it, the
result is a directional surprise, and reads as one. We do not claim
that pre-registration solves the simulator-validity problem; H4 FAIL
bounds those claims. Pre-registration solves the
\emph{within-simulation} interpretive problem, not the
\emph{simulation-to-reality} transfer problem.

\section{Limitations}
\label{sec:limitations}

We organize limitations as: pre-registered deviations, structural
constraints of the simulator, statistical-power constraints, and
transfer constraints.

\subsection{Pre-Registered Deviations}
Six deviations were logged during the campaign and disclosed in
\texttt{V40\_DEVIATIONS.md}~\cite{deviations_v40}. We summarize them
here.

\paragraph{DEV-1: H2 label swap.} The pre-registration uses
H2a $=$ blind and H2b $=$ cascade; the analysis code labels them
oppositely. The labels are cosmetic (both pairings are valid
paired-Wilcoxon tests of the same design), but the asymmetry between
the two CIs (H2b's wider CI under fidelity-filtered $n = 8$ vs
H2a's at $n = 9$) matters substantively. This paper uses
pre-reg labels throughout. A translation table is provided in the
artifact.

\paragraph{DEV-2: Pre-campaign artifact pollution.} Five
pre-campaign test runs without the campaign prefix were detected and
archived (not deleted) from the metrics directory before the primary
pipeline runs. No verdict effect.

\paragraph{DEV-3: C2 seed $= 18$ scenario-mole collision.} The
\texttt{enron\_blind.yaml} scenario contains a hardcoded
\texttt{fire: ALEKSANDER} event in Day 1 ($T_1$--$T_{21}$); seed
$= 18$ also assigns ALEKSANDER as the mole, which terminates the
mole agent before the pre-announce window opens. We exclude this run
from H1 and H2a ($n = 19$ each). A pre-flight scenario validator
asserting \texttt{mole\_identity $\notin$ scenario.fire\_targets} is
recommended for future versions.

\paragraph{DEV-4: H4 FAIL.} Pre-committed per pre-reg \S 5; see
\S\ref{sec:results-h4}.

\paragraph{DEV-5: Behavioral fidelity exclusions.} 23/40 adaptive
runs scored below the 3.5 PASS threshold. The exclusions are
measurement-artifact-driven (Spearman $\rho = 0.018$ for C3,
$-0.256$ for C4; neither significant) rather than signal-driven; see
\S\ref{sec:results-fidelity}. Per pre-reg \S 4.4, both primary
(fidelity-filtered) and sensitivity (full $N$) analyses are reported
side by side for H2.

\paragraph{DEV-6: Truncated runs.} Four runs (C2/1, C3/14, C4/7,
C5/9) terminated before $T_{100}$ due to infrastructure failures.
All four reached at least $T_{63}$, leaving the pre-announce
window and the $T_{60}$ in-degree snapshot intact. No verdict
changes.

\subsection{Structural Constraints}
\paragraph{Calibration (H4 FAIL).} HBEE's communication structure is
substantially more uniform than real organizations
(\S\ref{sec:results-h4}). The detection-inversion finding holds in
a stylized environment in which pair exposure is approximately
uniform. We do not claim transfer to environments with
Pareto-distributed exposure. The natural next step is a
heterogeneity-injected simulator (V41) calibrated to the SNAP
Enron Lorenz curve.

\paragraph{Single LLM backbone.} All agents are driven by a single
LLM (\S\ref{sec:design-llm}). The H5 cross-LLM robustness check is
PENDING (\S\ref{sec:results-h5}). Whether the inversion finding
replicates under a different model is an open question we leave to
future work.

\paragraph{Single mole archetype.} The OPSEC directive instantiates
the long-burn rationalizing-insider archetype (Snowden / Hanssen
pattern). Impulsive, financially motivated, and recruited-asset
archetypes have different behavioral signatures and may interact
differently with the cascade defender.

\paragraph{No multi-insider collusion.} HBEE assigns exactly one
mole per simulation. Multi-insider scenarios are operationally
relevant but introduce confounds we did not address.

\paragraph{Author-rated fidelity audit.} The fidelity audit is
single-rater. Inter-rater agreement is not estimated and could be
addressed in future work via blinded multi-rater protocols.

\subsection{Statistical-Power Constraints}
\paragraph{Small $N$.} $n_{\text{seeds}} = 20$ per condition is
adequate for the primary effects we observe (H1 narrowly via
sensitivity-bounded equivalence; H3 via large-effect Mann--Whitney)
but is small for sub-condition exploration. The H2b primary CI
illustrates the cost: at $n_{\text{pairs}} = 8$, the CI spans
$[-29.75, +14.25]$, wide enough that a substantively different
underlying effect could not be ruled out from this analysis alone.
The full-$N$ sensitivity analysis is the operative reading for H2b.

\paragraph{Single bootstrap RNG.} The 95\% CIs are computed from a
single bootstrap with seed $42$, $B = 1000$ resamples. Variation
across alternative RNG seeds is small in our spot checks but is not
formally characterized.

\subsection{Transfer Constraints}
\paragraph{Simulation-to-reality.} The detection inversion is a
result \emph{about HBEE}. Whether it transfers to real
organizations depends on whether the mechanism we propose (rate
reduction without signature change) generalizes. H4 bounds the
extent to which we can claim transfer. The most that can be
responsibly read from this paper is: \emph{in environments
structurally like HBEE, peer-suspicion-cascade detection inverts
under LLM-adaptive OPSEC}. The empirical question of whether real
organizations are structurally like HBEE in the relevant respects is
open.

\paragraph{Detector faithfulness.} The cascade and blind UEBA
mechanisms in HBEE are stylized implementations of the detection
families they represent. Real-world cascade detectors include
mechanisms (suspicion decay, anomaly weighting, multi-source
aggregation) that HBEE does not model. Our results are a starting
point for studying the family of mechanisms, not a verdict on any
specific deployed system.

\section{Ethics}
\label{sec:ethics}

\subsection{Dual-Use Considerations}
A paper demonstrating that an LLM-driven adaptive insider can evade
peer-suspicion-cascade detection is, on the surface, dual-use: a
defender can use the result to redesign cascade mechanisms; an
adversary can in principle use it to shape OPSEC behavior. We argue
the disclosure trade-off favors publication for three reasons.
First, the result is an empirical observation about a class of
detection mechanism, not an attack tool: there is no exploit code,
no specific evasion technique, and no guidance that does not already
appear in the operational-security literature accessible to any
motivated adversary. Second, the result is conditional on the
stylized environment HBEE provides; H4 FAIL bounds the
generalization claims explicitly. Third, defenders need this result
more than attackers do: peer-suspicion-cascade systems are
deployed; their fragility under adaptive adversaries is not
currently a recognized design constraint.

\subsection{IRB and Human Subjects}
This work involves no human subjects. The behavioral fidelity audit
rates LLM-generated outputs against an explicit Likert protocol; no
human participants are studied. No IRB review was sought because
none is applicable.

\subsection{Data Provenance}
HBEE uses the public SNAP Enron email graph~\cite{snap_enron} as a
calibration reference for the H4 Gini check. We use only the
graph's degree distribution; no message content, no individual
identities, and no personally identifiable information from the
Enron corpus are incorporated. The fictional Vixero HQ roster is
original; agent names, roles, and organizational structure are not
derived from any real organization.

\subsection{Deployment Risk}
A behavioral-only detection system trained on insights from this
work would carry a non-trivial false-positive risk: behavioral
patterns flagged as indicative of insider threat in a stylized
environment may correspond to ordinary employee experiences in a
real environment (medical issues, family circumstances, role
transitions, neurodivergence). The H3 detection inversion further
implies that any system designed to flag the most peer-suspected
agents may, in the presence of an adaptive adversary, flag innocents
disproportionately. Any operational deployment of insights from this
work should be a prioritization signal for human investigation, not
an automated adverse-action trigger. We discuss this further at
\cite{deployment_harm_addendum} in the artifact.

\subsection{Responsible Release}
We release the full simulator, frozen scenarios, raw telemetry, and
analysis code under an open-source license to support
reproducibility. The release does not include any production-system
configuration, real organizational data, or material that would
provide an adversary with capabilities not already publicly
accessible.

\section{Conclusion}
\label{sec:conclusion}

We presented HBEE, a 100-agent LLM-driven multi-agent simulator
designed for pre-registered hypothesis testing of insider-threat
detection mechanisms under adaptive adversary behavior. Across five
pre-registered conditions and 95 valid runs, we documented a
detection inversion: the adaptive mole's suspicion in-degree at
$T_{60}$ is statistically lower than that of a randomly selected
innocent agent (Cliff's $\delta = -0.694$, 95\% CI $[-0.855,
-0.519]$). A companion equivalence test (H2) shows that adaptive
OPSEC produces no detectable shift in the mole's UEBA rank under
either defender mode. Together, the two findings imply that the peer
suspicion graph and the per-agent UEBA score \emph{decouple} under
adaptive adversary behavior: the mole evades the social layer of
detection without moving the per-agent score that the social layer
is meant to amplify.

We bound transfer claims explicitly. A pre-registered Gini check
returns FAIL: HBEE's communication structure is substantially more
uniform than real organizations. The contribution of this paper is
narrow: a controlled, pre-registered demonstration of an unexpected
detection property in a stylized environment. We defend it as such.
Whether the inversion transfers to operational settings is an open
question that requires a calibrated successor system (V41) and
cross-LLM robustness work (H5) to resolve.

The paper's smaller methodological contribution is the
pre-registration discipline itself: a falsifiable hypothesis frame,
an explicit deviation log, equivalence-bound testing for null-shaped
claims, and a behavioral fidelity audit whose own structural
constraints became a finding. These practices may be useful to
others studying agent-based security phenomena where ``the
simulation behaved as expected'' is too easy to say in retrospect.

\section*{Acknowledgments}
The author thanks Amazon Web Services for compute resources used in the V40 campaign,
and the AgentSociety team (FIB Lab, Tsinghua University) for the open-source
\texttt{agentsociety} framework~\cite{piao2025_agentsociety} on which HBEE's
simulation infrastructure is built.

\section*{Open Science Statement}
The pre-registration document (\texttt{V40\_PREREG.md}, SHA256 hashed
prior to first run), full deviation log
(\texttt{V40\_DEVIATIONS.md}), simulator source, frozen scenario
YAMLs, raw JSONL telemetry for all 100 runs, behavioral fidelity
audit data, bootstrap-CI computation, completion table, and
figure-generation scripts are released at
\texttt{https://github.com/Vix0007/hbee-v40} under an open-source license. The
artifact is structured for USENIX artifact evaluation; a
\texttt{verify\_v40\_lock.sh} script regenerates SHA256 hashes for
the frozen artifacts.

\bibliographystyle{plain}
\bibliography{references}

\end{document}